\documentstyle[12pt,epsfig]{article}
\newcommand{\nc}{\newcommand}
 \nc{\be}{\begin{equation}}
 \newcommand{\ee}{\end{equation}}
 \nc{\beqa}{\begin{eqnarray}}
 \nc{\eeqa}{\end{eqnarray}}
 
\begin{document}
 \begin{titlepage}
 \pagestyle{empty}
 \baselineskip=18pt
 \rightline{BU-HEP 96-15}
 \baselineskip=21pt
 \vskip .2in
 \begin{center}
 {\large{\bf Transport Phenomena and Electroweak Baryogenesis\\
 in the Two-doublet Higgs Model}}
 \end{center}
 \vskip .2truecm
 \begin{center}
 Gian Franco Bonini

 {\it Physics Department\\
      Boston University\\
      Boston, MA 02215, USA
 }

 \end{center}

 \vskip 0.1in
 \centerline{ {\bf Abstract} }
 \baselineskip=18pt
 \vskip 0.5truecm
 We use  transport equations to compute the evolution of the quark
 plasma 
 at the electroweak phase transition in the two-doublet Higgs model,
 obtaining in a more rigorous and quantitatively accurate way results
 consistent with previous work. We discuss the
 model in connection with the electroweak baryogenesis scenario, and 
 claim that the observed baryon asymmetry
 of the universe can be obtained within this model with a suitable choice of parameters.

 \end{titlepage}

 \baselineskip=20pt
 \textheight8.3in\topmargin-0.0in\oddsidemargin-.0in

 \section{Introduction}
\setcounter{equation}{0}
 Electroweak baryogenesis was proposed years ago~{\cite{dimopoulos}}
 as a possible mechanism to generate the estimated
 baryon asymmetry of the universe ($n_{B}/s\sim 10^{-10}$), and it has been the 
 subject of a great
 amount of work ever since~{\cite{cohen93}}. Although the Standard Model at the
 electroweak phase transition satisfies the three necessary conditions
 pointed out by Sakharov~{\cite{sakharov}}, a quantitative
 analysis is far from being trivial. In principle, the actual computation
 of  the final baryon asymmetry requires a precise knowledge of the
 parameters (including those of the Higgs sector), of the details
 of the phase transition, of the $B$-violation rate near and at the phase
 transition, and of the transport properties of the QCD plasma in a
 non-uniform background field, which we do not yet have. A few
 general and persuasive conclusions have nonetheless been reached,
 and there is some consensus on the broad lines of this scenario:
 \begin{itemize}
 \item{At temperatures higher than the EW critical temperature, $B$
     violating processes~{\cite{tHooft}} are fast, and
  can wash out any previously
 created $B$ asymmetry. Soon below the critical
     temperature, B violation is heavily suppressed, and can be safely
     neglected, implying that the baryons we see now must have   
     been created at the EW phase
     transition. This argument assumes the absence of any primordial net $B-L$
     number, which no
   EW process   (including the anomalous ones)
     can wipe out and which would still be observable today, without any
                   need for baryon
     production at the weak scale. However,
in the absence of a compelling model
predicting such  a $B-L$ asymmetry,    we are going to
     ignore this second scenario. 
 }
 \item{A sufficiently strongly first-order PT is required, 
to prevent the newly
     created baryon asymmetry from being immediately washed out by the
     still strong $B$-violating processes.
 }

 \item{During nucleation, the expanding bubbles provide the necessary
     violation of $CP$ and (macroscopic) $CPT$, thereby creating a net
     amount of $CP$-odd fields. This process takes place at the
     bubble walls, where the variation in the background fields is
     most rapid.
 }

 \item{The actual $B$ generation takes place when the $CP$-odd fields 
       reach {\em{outside}} the bubble, where $B$-violating processes are
       fast. For this reason, it is important to understand the transport
       properties of those fields.
 }
 \item{In any case, the Minimal Standard Model is quantitatively
     inadequate because it does not
     contain enough $CP$ violation~{\cite{gavela}}, so that extensions to the MSM have to 
     be considered, and can be constrained (albeit loosely) by this
     cosmological test.
 }
 \end{itemize}

 In this paper, we focus on the transport properties of the plasma
 at the EW phase transition in the two-doublet Higgs model and discuss
 their effect on baryogenesis. In the next section we introduce the model
 and discuss its qualitative properties near the phase transition. In
 section 3 we show how to derive kinetic equations from microscopic
 principles (a result already present in the literature~{\cite{chou}}), 
 and  apply
 them to the problem at hand. Section 4 contains the explicit computation
 of the several parameters in the kinetic equations, and in section 5 we
 discuss our results and compare them with previous work.

 \section{The model}
\setcounter{equation}{0}
 As mentioned above, we need to extend the Minimal Standard Model to
 include more $CP$-violating terms. Our choice is to introduce a second
 Higgs doublet, with the following potential~{\cite{turok}}:
 \beqa
 \nonumber V(\phi_1,\phi_2)&=&\lambda_1(\phi_1^{\dagger}\phi_1-v_1^2)^2+
                  \lambda_2(\phi_2^{\dagger}\phi_2-v_2^2)^2+
                 \lambda_3[(\phi_1^{\dagger}\phi_1-v_1^2)\\
\label{pot}                       &+&(\phi_2^{\dagger}\phi_2-v_2^2)]^2+
                  \lambda_4[(\phi_1^{\dagger}\phi_1)(\phi_2^{\dagger}\phi_2)-
                            (\phi_1^{\dagger}\phi_2)(\phi_2^{\dagger}\phi_1)]\\
 \nonumber                 &+&\lambda_5[Re(\phi_1^{\dagger}\phi_2)-v_1v_2\cos\xi]^2+
                  \lambda_6[Im(\phi_1^{\dagger}\phi_2)-v_1v_2\sin\xi]^2
 \eeqa
 Yukawa interactions couple up-type quarks to $\phi_1$, and down-type
 quarks to either $\phi_1$ or $\phi_2$ (in order to avoid flavour
 changing neutral currents); it is irrelevant which one we choose,
 since we are going to neglect all couplings but the top's.
 The extra source of $CP$ violation is provided by the angle $\xi$, which
 cannot be rotated away unless $\lambda_5=\lambda_6$.

 What we are really interested in, however, is the effective action,
 which includes thermal and quantum corrections. Extremizing the
 effective action, we can obtain the profile of the bubble wall; this
 computation is discussed in~{\cite{cline}} 
 (at least in the case $\lambda_1=\lambda_2$,
 $v_1=v_2$, in which the two neutral Higgses undergo a phase transition
 simultaneously). For the moment, however, we do not need an explicit
 solution, and  write the VEV's in the general form:
 \be
 <\phi_i^0(x)>=\rho_i(x)\exp(-i\theta_i(x))
 \ee
 In the following, we shall use $h_i^0$, $h_i^{-}$ ($i=1,2$) to refer to
 the four (complex) excitations in the Higgs fields.

 An important quantity here is the width of the wall compared to the
 particle mean free path; the latter can be estimated using the cross
 section of the typical QCD process at this energy scale,
 i.e. gluon-mediated quark-quark scattering: as we shall show later,
  $\Gamma_{QCD}\sim 
 T/3$. On the other hand, the wall width is strongly model-dependent, and
 even in the specific theory we are discussing it has not yet been
 computed precisely. For this reason, it would be useful to develop a
 formalism able to deal with both the ``thin wall'' and the ``thick
 wall'' limits, as well as the intermediate regime. Attempts in this
 direction will be discussed in the last section. Our work here applies,
 at least in its simplest form, to the thick wall limit, where the
 particles inside the wall can be taken to be in local kinetic
 equilibrium.
 Under those conditions, it is expedient to follow~{\cite{cohen336}} 
 and perform a
 hypercharge rotation of the fields:
 \be
 f_i\rightarrow e^{2iy_i\theta_1(x)}f_i
 \ee 
 where $f_i$ is a generic field, and $y_i$ its hypercharge.

 After the rotation, the spacetime-dependent top mass becomes real (as we
 said, we are neglecting all other masses), but a new interaction
 term appears:
 \be
\label{int}
 {\cal L}_{hyp}=-2\partial^{\mu}\theta J^Y_{\mu}
 \ee
 where $J^Y_{\mu}$ is the hypercharge current.
 
Our task is to study the effect of this term on the plasma density; 
 let's first recapitulate the essential features of the
 system:

 \begin{itemize}
 \item{The EW phase transition takes place at a temperature $T_c\sim 100
     GeV$. At this energy scale quarks and gluons are deconfined, and the
     strong coupling constant is small enough ($\alpha_s \sim 0.1$) for
     perturbation theory to be reliable. 
 }
 \item{The Hubble expansion rate ($H\sim T^2/m_{pl}$) is much smaller than
     any other interaction rate, and can be ignored.
 }
 \item{We shall assume that each particle species $i$ is locally in kinetic
     equilibrium, so that its density is determined by its chemical potential
     $\mu_i(x)$ and temperature.
 }
 \item{In principle, baryon production depends on the densities of all
     of the left-handed baryons
     and leptons. However, we shall ignore all leptons and the two lighter
     quark families,
     since their Yukawa couplings are much smaller than the top's.
 }
 \end{itemize}

 With those assumptions, the system can be described in terms of just
 three quantum number densities:
 \beqa
 \label{q} q(x)&\equiv& t_L(x)+b_L(x)\\
 t(x)&\equiv& t_R(x)\\
 \label{h} h(x)&\equiv& h_1^0(x)+h_1^{-}(x)+h_2^0(x)+h_2^{-}(x)
 \eeqa

Below, we shall derive evolution ({\it i.e.}, transport) equations for the quantities
(\ref{q})-(\ref{h}), in the presence of the non-uniform external
interaction (\ref{int}). Unlike \cite{cohen336}, which addressed the same
problem, we  compute the relevant parameters in terms of
microscopic physics, using a general technique that can be applied to 
other (possibly more realistic) extensions of the Standard Model.

 \section{Transport equations}
\setcounter{equation}{0}
 We refer to~{\cite {chou}} for a derivation of macroscopic transport
 equations from microscopic principles;
 here we limit ourselves to presenting the final result and discussing its
 physical significance.

 The key idea is that for quasiuniform systems (such as ours), it is
 possible to define local densities $n_i({\bf k},x^{\mu})$ 
 of  particles  with given on-shell
 momentum. The time evolution of these quantities can be derived by computing
 appropriate Green functions ({\it i.e.} Feynman diagrams) involving the
 elementary fields.

 In the simple case of a scalar field with lagrangian: 
 \be
 {\cal
   L}=\frac{1}{2}\partial_{\mu}\phi(x)\partial^{\mu}\phi(x)-
 \frac{1}{2}m^2\phi^2(x)-V(\phi(x))
 \ee
 we obtain 
 \be{\label{transport}}
 \frac{\partial n({\bf k},x^{\mu})}{\partial t}+{\bf v}\cdot{\bf \nabla}
 n({\bf k},x^{\mu})+\frac{\partial
   \omega}{\partial x_{\mu}}\frac{\partial n({\bf k},x^{\mu})}{\partial 
 k^{\mu}}=W_e(1+n({\bf k},x^{\mu}))-W_an({\bf k},x^{\mu})
 \ee
 where  $\omega$ is given by the space-time dependent dispersion
 relation, and $W_e$ and $W_a$ are the emission and absorption rates:
 \beqa
 W_a({\bf k})&=&\frac{1}{2\omega}\sum_{l,n}\left|<l|j(0)|n>
 \right|^2\rho_{nn}(2\pi)^4\delta^4(k-p_l+p_n)\\
 W_e({\bf k})&=&\frac{1}{2\omega}\sum_{l,n}\left|<n|j(0)|l>
 \right|^2\rho_{ll}(2\pi)^4\delta^4(k-p_l+p_n)\\
 j(x)&\equiv&-\frac{\delta V(\phi(x))}{\delta \phi(x)}
 \eeqa
Here $\rho$ is the density matrix of the system, and the sums run over a
 complete set of states of the system.
 Eq.~({\ref{transport}}) applies to quasiuniform systems,
  but does not rely on the coupling constants
 being small, so that in principle we can compute $W_a$ and $W_e$ to any 
 order in perturbation theory, provided that we use the correct
 finite-temperature Feynman rules (for a thorough discussion of the
 finite-temperature formalism, see
 {\cite{landsman}}).

This
   procedure can be easily extended to more complicated cases, and  to
   fermions; in the latter case, the statistical factors of $(1+n)$ 
    become $(1-n)$. For each particle species we obtain an equation of
    the form (\ref{transport}) and every scattering process  contributes
    a term to the $W_e$ ($W_a$) of each emitted (absorbed) field. 

 In the case of kinetically thermalized particles, the  reaction channel
 $A_1...A_n\rightarrow B_1...B_m\phi$ contributes to  
 the functions $W_a$ and $W_e$ corresponding to $\phi$ as follows:
 \beqa
 W_a({\bf k})&=&\frac{1}{2\omega}\int
 \prod _{A,B}\frac{d^3p_i}{(2\pi )^3 2E_i}
 \left| j_{B\phi\rightarrow A} \right|^2
 (2\pi )^4\delta^4(k+P_{B}-P_{A})
 \prod _{B}n_i\prod _{A}(1+n_i)\\
 W_e({\bf k})&=&\frac{1}{2\omega}\int
 \prod _{A.B}\frac{d^3p_i}{(2\pi )^3 2E_i}
 \left| j_{A\rightarrow B\phi} \right|^2
 (2\pi )^4\delta^4(k-P_{A}+P_{B})
 \prod _{A}n_i\prod _{B}(1+n_i)
 \eeqa
 so that the r.h.s. of Eq.~(\ref{transport}) becomes:
 \be
 \frac{1}{2\omega}\int
 \prod _{A,B}\frac{d^3p_i}{(2\pi )^3 2E_i}
 \left| j_{B\phi\rightarrow A} \right|^2
 (2\pi )^4\delta^4(k+P_{B}-P_{A})
 [n_{\phi}\prod _{B}n_i\prod _{A}(1+n_i)-
 (1+n_{\phi})\prod _{A}n_i\prod _{B}(1+n_i)
 ]
 \ee
 We thus recover the usual detailed balance conditions, and the (local)
 equilibrium density of a given particle species depends only 
 on its dispersion relation.
 
 Therefore, given a space-time dependent 
 perturbation, the behaviour of the system is affected in two ways:
 \begin{itemize}
 \item{Dispersion relations are modified, and that changes the local
     equilibrium densities;}
 \item{Source terms $W_e$ and $W_a$ appear, and they determine the
     rate at which local equilibrium is approached.}
 \end{itemize}

 In order to find the total number density, we integrate 
 Eq.~({\ref{transport}}) over ${\bf k}$. Elastic reactions (which do not
 change the identity of the involved particles, but only their momenta)
 give rise, after the integration is performed, to the usual diffusive
 behaviour. The interactions that do violate some particle quantum number,
 on the other hand, lead to net source terms (they contribute to
 diffusion as well, but we shall ignore this effect, because of its
 smallness).

 In the next section, we shall apply Eq.~({\ref{transport}}) to our problem,
  considering both effects in detail, and
 presenting some explicit computations.

 \section{Results}
\setcounter{equation}{0}
 The dispersion relation of a particle $i$ is affected by the interaction
 ${\cal L}_{hyp}$ as follows:

 \be
 k^{\mu}k_{\mu}=m_i^2 \rightarrow (k^{\mu}+2y_i\partial^{\mu}\theta)^2=m_i^2
 \ee
 As mentioned above, this
   automatically provides  values for the local equilibrium
 densities~{\cite{cohen336}}:

 \be \label{equil}
 f_{i,eq}=\frac{K_i\bar \mu_iT^2}{6}
 \ee
 where $K_i=$ (number of spin degrees of freedom)$\times 2(1)$ for bosons
 (fermions), and $\bar\mu_i=-2\dot\theta y_i$.

 Assuming kinetic equilibrium, the local densities are of the form
 \be
 n_i({\bf k})=\frac{e^{-\beta(E_i({\bf
       k})-\mu_i)}}{1\mp e^{-\beta(E_i({\bf k})-\mu_i)}}
 \ee

 where $E_i=\sqrt{{\bf k}^2+m_i^2}-\bar\mu_i$.

 In the thick wall approximation, $\theta$ varies slowly, so that we can ignore the
 third term on the l.h.s. of Eq.~({\ref{transport}}), which involves
 the {\em second} derivatives of $\theta$, and we only  need  compute the
 source terms $W_e$ and $W_a$.

 \begin{itemize}
 \item{
 {\bf Quark elastic scattering} is dominated by QCD processes (see
 Fig.{\ref{fig1}}).
\begin{figure}[ht]
\epsfig{file=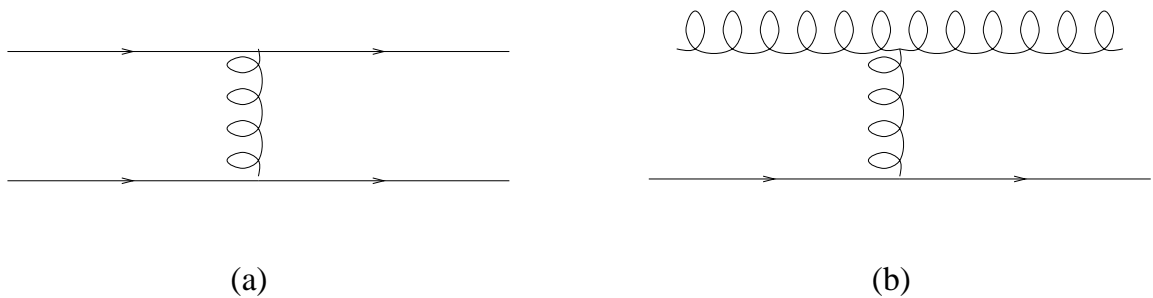,width=1.5truein,height=6truein, angle=270}

 \caption{Dominant processes responsible for quark diffusion.}
 \label{fig1}
 \end{figure}

 To lowest order, these processes can be computed 
 assuming a perfectly thermalized system. In this case we do not have to
 use Eq.~({\ref{transport}}) and can instead apply standard finite-temperature
 field theory~{\cite{landsman}}.

 The finite-temperature rate of quark-scattering processes  
 can be obtained (via the
 optical theorem) from the imaginary part of the quark self-energy. This
 computation is discussed in~{\cite{lebedev}}; as  emphasized there,
 diffusion depends not on the {\em total} cross  section $\int 
 d\sigma/d\Omega$, but on the {\em transport} cross section 
 $\frac{3}{2}\int   d\sigma/d\Omega \sin^2\theta$, where $\theta$ is
 the scattering angle (qualitatively, this is because close-to-forward
 scattering does not contribute much to diffusion).  Taking that into
 account, we obtain  the interaction rate for a particle with
 momentum ${\bf p}$:\footnote{$Im \Sigma^R$
   in~{\cite{lebedev}} is related to $W_e$ and $W_a$ by:
 \beqa \nonumber Im \Sigma^R&=&-\Gamma\gamma^0/2; \\
 \nonumber
 \Gamma&=&W_e(1+e^{\beta E})\\
 \nonumber
      &=&W_a (1+e^{-\beta E})
 \eeqa}  
 \be
 \Gamma^{diff}_{quark}({\bf p})=\frac{9}{32\pi }\omega_{pl}^2g^2TC_F
 \int_{-1}^1d\cos\theta\int
 \frac{dk^2}{k^4+\frac{9}{16}\omega_{pl}^4\cos^2\theta}
 \frac{k^2\sin^2\theta}
 {p^2+k^2-2pk\cos\theta}
 \ee

 where $p=|{\bf p}|$, $C_F=4/3$, and $\omega_{pl}^2=(C_V+N_f/2)g^2T^2/9$, 
 with $C_V=3$ and $N_f=6$.

 The leading contribution comes from the infrared behaviour of the
 integrand, so we can write:

 \be
 \Gamma^{diff}_{quark}({\bf p})\sim\frac{9}{32\pi }\omega_{pl}^2g^2TC_F
 \int_{-1}^1d\cos\theta\frac{1-\cos^2\theta}{p^2}\int
 \frac{dk^4}{2(k^4+\frac{9}{16}\omega_{pl}^4\cos^2\theta)}
 \ee

 and, in the leading logarithm approximation,

 \be
 \label{diff_quark}
 \Gamma^{diff}_{quark}({\bf
   p})=\frac{2\pi }{3}\frac{T^3}{p^2}
 C_F(C_V+\frac{N_f}{2})\alpha^2_s\ln\alpha_s^{-1}\sim 0.4 \frac{T^3}{p^2}
 \sim\frac{T}{3}
 \ee
 (the last step was obtained by averaging over $p$).
 }

 \item{

 {\bf Higgs elastic scattering} is dominated by electroweak processes
 (Fig.{\ref{fig2}}):

 \begin{figure}[ht]
\epsfig{file=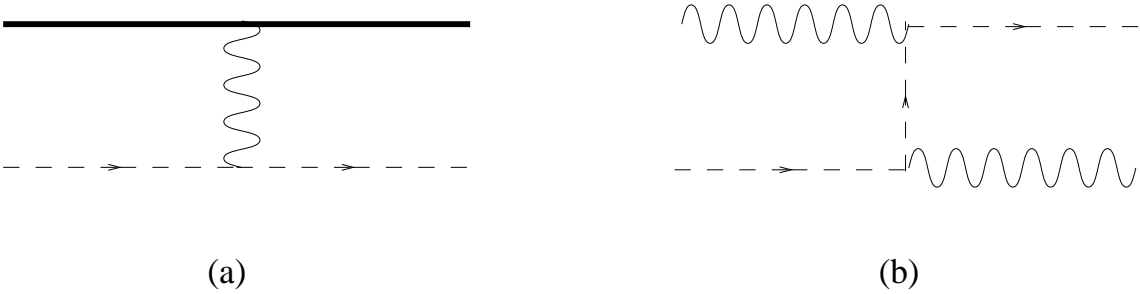,width=1.5truein,height=6truein, angle=270}
 \caption{Dominant processes responsible for Higgs diffusion (wiggly
   lines stand for electroweak bosons, dashed lines for Higgses, and
   double lines for any particle coupled to electroweak bosons).}
 \label{fig2}
 \end{figure}

 Diagram 2a is similar to those computed above, except that
 all of the particles here are bosons. A further difference is due to the
 vector boson mass, which provides an IR cutoff: the W and  Z bosons
 receive mass not only from quantum corrections, but also from the
 coupling with the Higgs VEVs. However, both contributions to $m^2_{W,Z}$
 are proportional to $g^2_{W}T^2$, and in the leading log
 approximation the precise coefficient is irrelevant. The $SU(2)_L$ and
 the $U(1)_Y$ sectors each contribute  with a term:
 \be
 \label{diff_Higgs}
 \Gamma^{diff}_{Higgs}({\bf
   p})=\frac{4\pi }{3}\frac{T^3}{p^2}
 C_F(C_V+\frac{N_f}{2})\alpha^2\ln\alpha^{-1}
 \ee

 In the $SU(2)_L$ term $C_V=\frac{3}{4}$, $C_F=2$, $\alpha=g^2/4\pi $ and
 $N_f$ is the number of doublets coupled to the bosons; 
 in the $U(1)_Y$ term $C_V=0$, $C_F=1$, $\alpha=g'^2/4\pi $ and
 $N_f=2\sum y_i^2$, summed over all the hypercharged particles. 
  The total diffusion width from Diagram 2a is 
 \be
 \Gamma^{diff}_{Higgs}({\bf
   p}) \sim 0.15 T
 \ee

 The Higgs being a boson, Diagram 2b has apparently the same IR behaviour
 as Diagram 2a, and has to be considered as well;
 however, in this case the exchanged particle has a mass $\sim O(\lambda
 T)$, where $\lambda$ is the typical Higgs self-coupling constant (see
 Eq.~(\ref{pot})); if $\lambda \sim 1$, 
then  the diagram contributes a subleading $\alpha^2$ term, which we neglect.
 }

 Both for quarks and for Higgses, the diffusion constant is related to $\Gamma$ by the
 usual formula:

 \be
 D=\frac{1}{3\Gamma^{diff}}
 \ee

 \item{
 {\bf top-quark mass coupling}, which mixes $q$ and $t$ (Fig.3a):

 \begin{figure}[ht]
\epsfig{file=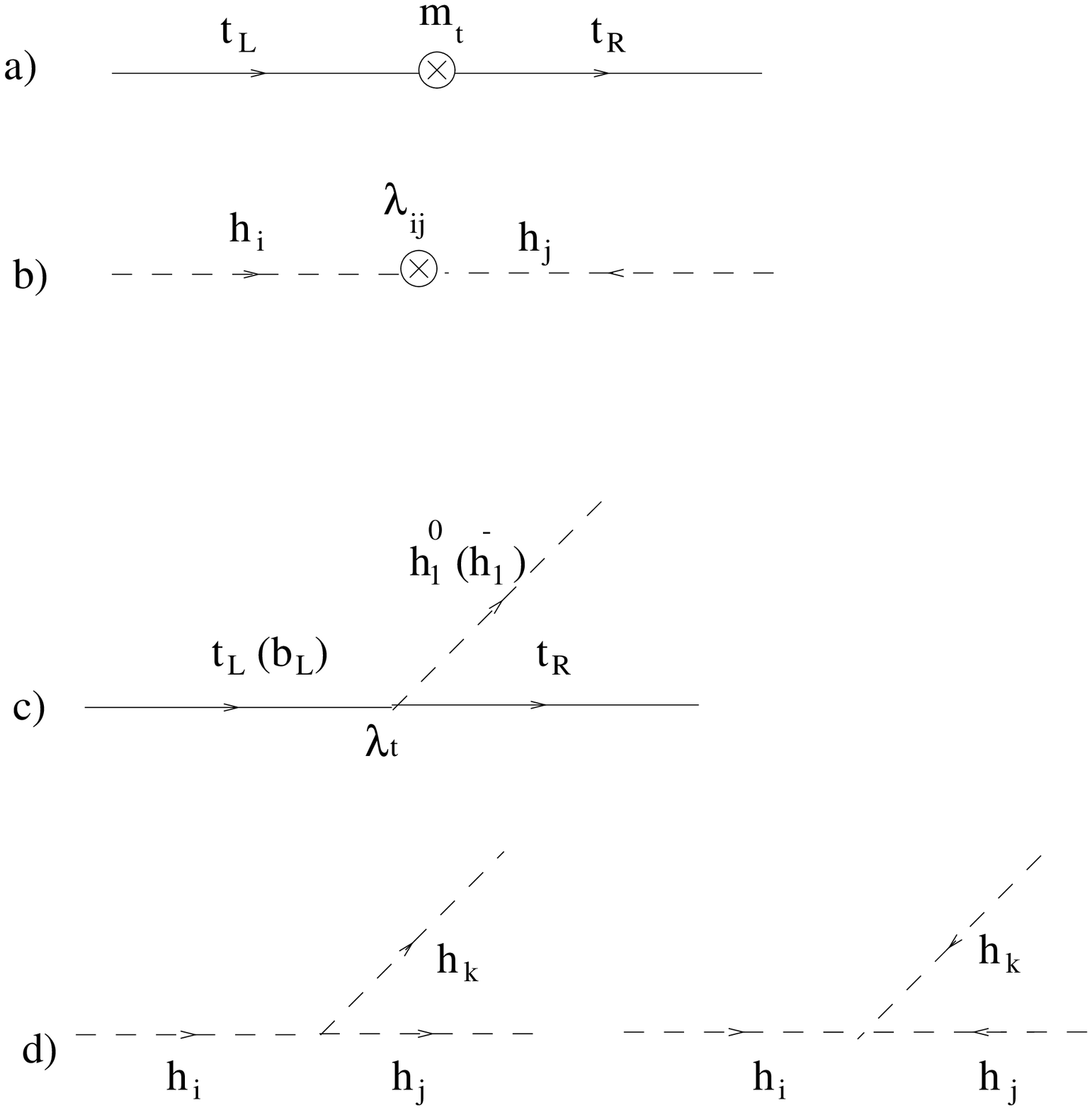,width=6truein,height=6truein}
 \caption{Vertices that violate at least one among quantum numbers $t$,
   $q$, $h$.}
 \label{fig3}
 \end{figure}
 Strictly speaking, we
 cannot compute an interaction {\em rate} associated to such diagrams,
 because a quadratic perturbation leads to particle
 oscillations, not to exponential decay. However, it is still possible to
 compute an average rate
 (which does not vanish, since the particles experience thermal damping as
 well) as follows:
 \begin{itemize}
 \item{
 As described in~{\cite{lebedev}}, the top quark retarded propagator in
 momentum space, including thermal correction, has the
 form:
 \be
 G(p_{\mu})=\frac{i\left[ \left( p_0-\Sigma_0\right)\gamma^0-{\bf
       p}\cdot {\bf \gamma}\left( 1-\frac{\Sigma_3}{
         p}\right)\right]}{\left( p_0-\Sigma_0\right)^2 -p^2\left( 1-\frac{\Sigma_3}{p}\right)^2}
 \ee
 where $p=|{\bf p}|$ and $\Sigma_0$ and $\Sigma_3$ are functions of $p_0$
 and $p$. The formula above is valid for massless fermions, and does not
 mix helicities.
 }
 \item{
 Adding the top mass introduces an off-diagonal term in the helicity
 basis:
 \be
 G_{LL}=G_{RR}=\frac{i\left[ \left( p_0-\Sigma_0\right)\gamma^0-{\bf
       p}\cdot {\bf \gamma}\left( 1-\frac{\Sigma_3}{
         p}\right)\right]}{\left( p_0-\Sigma_0\right)^2 -p^2\left( 1-\frac{\Sigma_3}{p}\right)^2-m_t^2}
 \ee
 and
 \be
 G_{LR}=G_{RL}=\frac{-im_t}
         {\left( p_0-\Sigma_0\right)^2 -p^2\left( 1-\frac{\Sigma_3}{p}\right)^2-m_t^2}
 \ee
 }
 \item{
 We now Fourier transform $G_{LR}$ with respect to time and obtain:
 \be
 G_{LR}(t,{\bf p})=-\frac{im_t}{\omega_p}e^{-\Gamma_b t/2}\sin(\omega_pt)
 \ee
 where $\omega_p^2=p^2(1-\Sigma_3/p)^2=p^2+g_s^2T^2C_F/4$, and
 $\Gamma_b=2Im
 \Sigma_0=\alpha_s\pi ^2C_FT/6\log(\alpha_s^{-1})\sim0.5 T$ 
 is the total scattering rate with the thermal bath. 
 }
 \item{we can then compute the probability of finding a right top $t_R$ at time
     $t+\Delta t$ coming from  a left top  at time $t$:
 \be
 P(R\rightarrow L)(\Delta t)=\left|G_{LR}(\Delta t)\right|^2=
 \frac{m_t^2}{\omega_p^2}e^{-\Gamma_b\Delta t}\sin^2(\omega_p \Delta t)
 \ee
 }
 \item{finally, we can compute the rate of change of the right and left
     top quantum numbers; this is the sum of two terms, since to the
     time derivative of $P(R\rightarrow L)$ we must add the decay
     products of the left top itself, which have the same quantum number.
 We therefore obtain:

 \beqa
\nonumber
 W_e(t_L\rightarrow t_R)&=&\frac{m_t^2}{\omega_p^2}
 \frac{\int_0^{\infty}dt\;\left(\frac{d}{dt}\left(e^{-\Gamma_b t}
 \sin^2(\omega_p t)\right)+\Gamma_b\left(e^{-\Gamma_b t}\
 sin^2(\omega_p t)\right)\right)}{\int_0^{\infty}dt\;e^{-\Gamma_b t}}t_L\\
           &=&\frac{2\Gamma_bm_t^2}{\Gamma_b^2+4\omega_p^2}t_L
 \eeqa
 and, using Eq.~({\ref{transport}}):
 \be
 \dot t_R({\bf p})
   =\frac{2m_t^2\Gamma_b}{(\Gamma_b^2+4\omega_p^2)}(t_L({\bf p})-t_R({\bf p})) 
 \ee

 In the limit of large $\Gamma_b$, this rate vanishes. As mentioned
 in~{\cite{nelson}}, this happens because thermal interactions destroy
 quantum coherence, which is a necessary condition for these processes
 to occur.

 Integrating over ${\bf p}$ we finally obtain, in terms of variables (7)-(9):
 \be
 \label{gamma_mass}
 \dot t=\frac{m_t^2\Gamma_b}{32\pi ^2T^2}(q-q_{eq}-2(t-t_{eq}))\equiv
 \Gamma_{mass} \left(\frac{q}{6}-\frac{t}{3}-\frac{h_{eq}}{8}\right)
 \ee
 where $q_{eq}$, 
 $t_{eq}$, and $h_{eq}$ have been defined in Eq.~(\ref{equil}).
 }
 \end{itemize}
 }
 \item{
 {\bf Higgs mass matrix} (Fig.3b):
 the Higgs mass matrix should in principle be obtained from the quadratic
 term in the expansion of the effective action around the bubble
 solution; both outside and deep inside the bubble, the Higgs VEVs are
 slowly varying, and we can obtain the Higgs masses by expanding the
 finite temperature effective {\em potential}, $V_T(\phi)$ (we assume
 that the transition is first order, so in both regions $V_T(\phi)$ is convex
 and the masses are well defined). In the bubble wall, where the field
 configuration is more strongly space-time dependent, we have to expand
 the effective {\em action}. Unfortunately, we do not have an explicit
 expression for $V_T(\phi)$, nor for the effective action, so we shall
 just assume that the mass matrix elements are of order $T$.
 Higgs number violation is due to 
 terms proportional to $h_ih_j$ or $ h_i^{\dagger}  h_j^{\dagger}$, which 
 are responsible for two point Green
 functions $G_{h_ih_j}$.
 Explicitly, the quadratic part of the neutral \footnote{The electric
   charge is not spontaneously broken, so terms of the form
   $h_i^{-}h_j^{-}$ do not appear, and we can disregard charged Higgses altogether.}
 Higgs self-coupling can be written (in Fourier space) as:
 \be
 {\cal L}^{(2)}=\frac{1}{2}H^{\dagger}({\bf k})
 \left(\begin{array}{cc}k^2-M&\Lambda\\
                        \Lambda^{\dagger}&k^2-M\\
 \end{array}
 \right)H({\bf k})
 \ee

 where 
 \be
 H=\left(\begin{array}{rc}
 h_1^{0}\\
 h_2^{0}\\
 h_1^{0\dagger}\\
 h_2^{0\dagger}\\
 \end{array}
 \right)
 \ee
 and $M$ and $\Lambda$ are $2\times 2$-matrices. 

 Thermal interactions modify the dispersion relations by adding
 ``mass-like'' terms that are much smaller than the original masses, so
 we neglect this effect;  thermal damping is taken into account by
 letting
 \be
 k_0\rightarrow \tilde k_0 \equiv k_0 + i\Gamma/2
 \ee
 where $\Gamma$ is the total Higgs scattering rate computed above.

 $\Lambda$, which is
 responsible for Higgs number violation, will be taken to be small (in
 the opposite regime, we could content ourselves with the local
 equilibrium approximation); under that assumption, and in the basis
 that diagonalizes $M$, the
 propagators are:
 \be
 |G_{h_ih_j}|^2=e^{-\Gamma t}\sin^2\left(\frac{|\lambda_{ii}|t}{2\sqrt{{\bf k}^2+m_i^2}}\right)\delta_{ij}
 \ee
 and the same procedure as above yields to:
 \be
 \dot h=-\Gamma_{Higgs}^{(1)}\frac{h-h_{eq}}{8}
 \ee
 where
 \be
 \label{gamma_Higgs}
 \Gamma_{Higgs}^{(1)}= \frac{\Pi ^2(|\lambda_{11}|+|\lambda_{22}|)}{4\zeta(3)\Gamma}
 \ee
 }
 \item{
 {\bf Higgs Yukawa couplings} (Fig.3c):

 We shall assume that the Higgs that couples to the top is sufficiently light
 ($m_h<m_t-m_b $) or heavy ($m_h>m_t+m_b $), so that either 
 $t_R\rightarrow b_Lh^{+}$ or $h^{-}\rightarrow b_L\bar t_R$ can occur on shell;
 then, the leading contribution to $|j|^2$ is simply $\lambda_t^2$, with
 a normalization factor of $2m_i$ for each fermion.

 In the first case, 
 we obtain:
 \beqa
 \dot h&=&2\lambda_t^2\int \frac{d{\bf k}_t}{(2\pi )^32E_t} \frac{d{\bf
     k}_h}{(2\pi )^3 2E_h}
  \frac{d{\bf k}_b}{(2\pi )^32E_b}\cdot 4m_t m_b
  (2\pi )^3\delta^3({\bf k}_h+{\bf k}_b-{\bf k}_t) \\ \nonumber &&2\pi \delta\left(
    E_h-E_t+\sqrt{m_b^2+
\left({\bf k}_h-{\bf k}_t\right)^2
}\right)[
 (n_h+1)(1-n_b)n_t-n_hn_b(1-n_t)]
 \eeqa

 As we said, we assume kinetic equilibrium distributions, so:
 \be
 \dot
 h=\frac{\lambda_t^2m_tm_b}{4\pi ^3}\int_{m_t}^{\infty}dE_t
 \int_{E_m}^{E_M}dE_h\frac{e^{-\beta E_t}\beta
   (-\mu_t-\mu_h+\mu_b)} {(1-e^{-\beta(E_h+\mu_h)})
 (1+e^{-\beta(E_b-\mu_b)})(1+e^{-\beta(E_t-\mu_t)})}
 \ee

 where $E_m$ and $E_M$ are purely determined by the kinematics:
 \be
 E_m=E_t\frac{m_t^2+m_h^2-m_b^2}{2m_t^2}-\sqrt{E_t^2-m_t^2}
 \frac{\sqrt{(m_t^2+m_h^2-m_b^2)^2-4m_t^2m_h^2}}{2m_t^2}
 \ee
 \be
 E_M=E_t\frac{m_t^2+m_h^2-m_b^2}{2m_t^2}+\sqrt{E_t^2-m_t^2}
 \frac{\sqrt{(m_t^2+m_h^2-m_b^2)^2-4m_t^2m_h^2}}{2m_t^2}
 \ee
 (in our case, we can safely ignore $m_b$, which is much smaller than the other masses)

 Therefore:

 \be {\label{yuk}}
 \dot h=\dot t=-\dot
 b=-\frac{3\lambda_t^2m_tm_b}{2\pi ^3T}\left(\frac{h}{8}-\frac{q}{6}
 +\frac{t}{3}\right)A\equiv -\Gamma_{yukawa}\left(\frac{h}{8}-\frac{q}{6}
 +\frac{t}{3}\right)
 \ee
 where
 \be
 A(t\rightarrow bh^{+})=\beta\int_{m_t}^{\infty}dE \frac{e^{\beta E}}{(1+
 e^{\beta E})^2}\log\left(\frac{e^{\beta E_M}-1}{e^{\beta E_m}-1}
 \frac{e^{\beta E_m}+e^{\beta E}}
 {e^{\beta E_M}+e^{\beta E}}\right)
 \ee

 is of order 1.

 In the second case, Eq.~({\ref{yuk}}) still applies, with

 \be \label{hbt}
 A(h^{-}\rightarrow b\bar t)=\beta \int_{m_h}^{\infty}dE \frac{e^{\beta E}}{(1-
 e^{\beta E})^2}\log\left(\frac{e^{\beta E_M}+1}{e^{\beta
     E_m}+1}\frac{e^{\beta E_m}+e^{\beta E}}
 {e^{\beta E_M}+e^{\beta E}}\right)
 \ee

 and the roles of $m_t$ and $m_h$ in $E_M$ and $E_m$ are interchanged.

 If $h^{0}\rightarrow t\bar t$ is also allowed  (which is true at least 
 in some region of the bubble wall,
 where the Higgs VEV, and consequently $m_t$, are sufficiently small),
 then we have another term, larger than (\ref{hbt}) by a factor of
 $m_t/m_b$, which therefore dominates.

 In the narrow window $m_t-m_b<m_h<m_t+m_b$, 
 all these  processes are kinematically forbidden and  we must go to higher
 order, including QCD corrections, like the one shown in Fig.{\ref{fig4}}.
 The evaluation of this graph proceeds as in the previous computations;
 the exchanged particle is a fermion, so that both its propagator and its
 statistical factor make the diagram less IR divergent; the IR cutoff
 $\lambda$ is
 provided by whichever is largest among the zero-temperature quark
 masses, their thermal corrections, and $|m_h-m_t|$. In any case, the
 result is suppressed, with respect to the previous case, by $\alpha_s$:

 \be
 \Gamma_{yukawa} \sim \frac{\lambda_t^2\omega_{pl}|m_h-m_t|}{\Lambda T}
 \ee

 \begin{figure}[ht]
\begin{center}
\epsfig{file=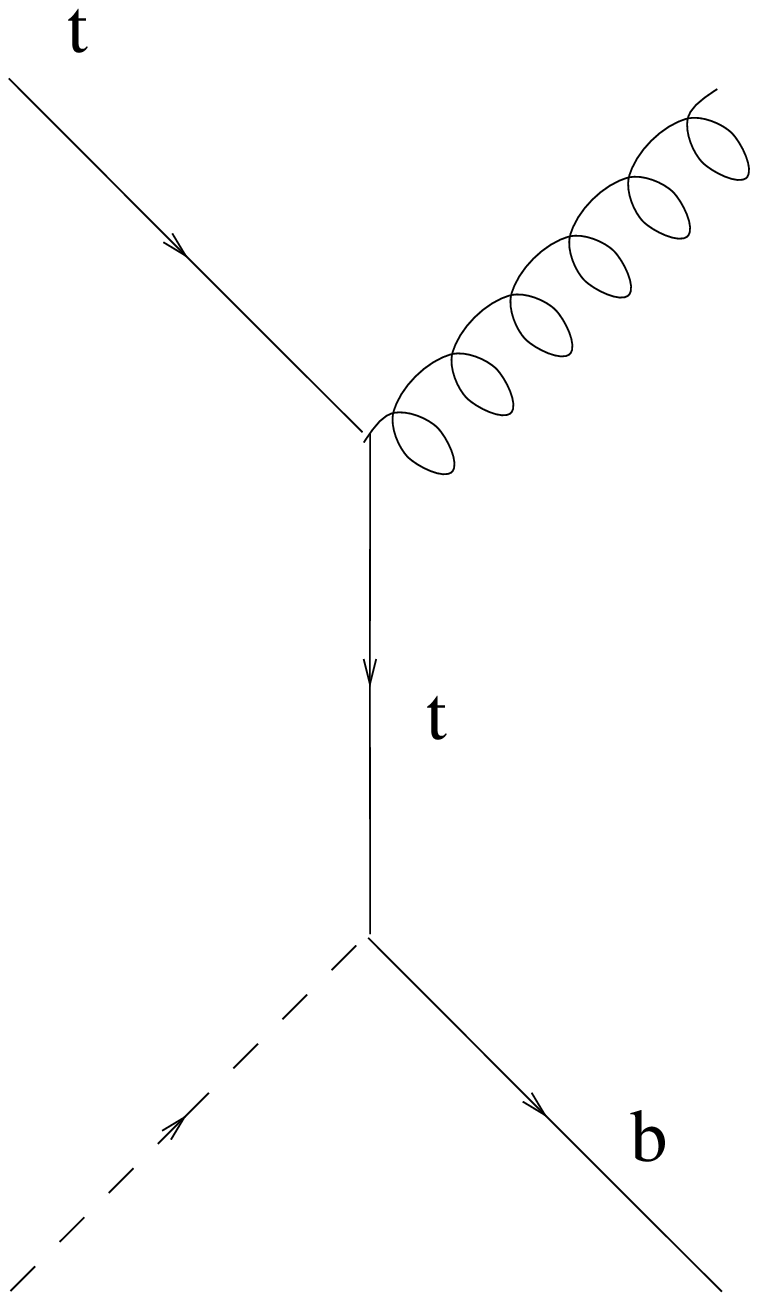,width=2truein,height=2truein}
 \caption{}
\end{center}
 \label{fig4}
 \end{figure}
}
 \item{
 {\bf Higgs cubic coupling}:
 the effects of cubic Higgs couplings (Fig3d)  depend on the Higgs 
 masses; if at least one process of the form
 $h_i\rightarrow h_jh_k$ can occur on shell, its contribution can be
 evaluated as above (with the small modifications due to the different statistics):

 \beqa
 \dot h&=&\sum_{ijk}2|\lambda_{ijk}|^2\int \frac{d{\bf k}_i}{(2\pi )^32E_i} \frac{d{\bf
     k}_j}{(2\pi )^3 2E_j}
  \frac{d{\bf k}_k}{(2\pi )^32E_k}
  (2\pi )^3\delta^3({\bf k}_j+{\bf k}_k-{\bf k}_i)\cdot \\
\nonumber &&2\pi \delta\left(
    E_j-E_i+\sqrt{m_k^2+
({\bf k}_i-{\bf k}_j)^2}\right)[
 (n_j+1)(1+n_k)n_i-n_jn_k(1+n_i)]
 \eeqa
 here the sum includes all combinations of Higgs and anti-Higgs fields
 that are kinematically allowed, and 
 $\lambda_{ijk}$ is a function of $\lambda_1$...$\lambda_6$ and of the Higgs
 VEVs.}

 We therefore obtain:

 \be
 \label{gamma_Higgs_2}
 \dot h=-\sum_{ijk}\frac{3|\lambda_{ijk}|^2}{64\pi ^2T}
 \int_{m_i}^{\infty}dE \frac{e^{-\beta E}}{1-
 (e^{-\beta E})^2}\log\left(\frac{e^{\beta E_M}-1}{e^{\beta
     E_m}-1}\frac{e^{\beta E_m}-e^{\beta E}}
 {e^{\beta E_M}-e^{\beta E}}\right)(h-h_{eq})\equiv
 -\Gamma_{Higgs}^{(2)}\frac{h-h_{eq}}{8}
 \ee

 The integral, formally divergent as $m_i \rightarrow 0$, is of order $T$
 if the Higgs masses are $\sim T$, as we expect.

 If no 3-Higgs process is kinematically allowed, electroweak corrections
 must be included (Fig.\ref{fig5}).
 \begin{figure}[ht]
\begin{center}
\epsfig{file=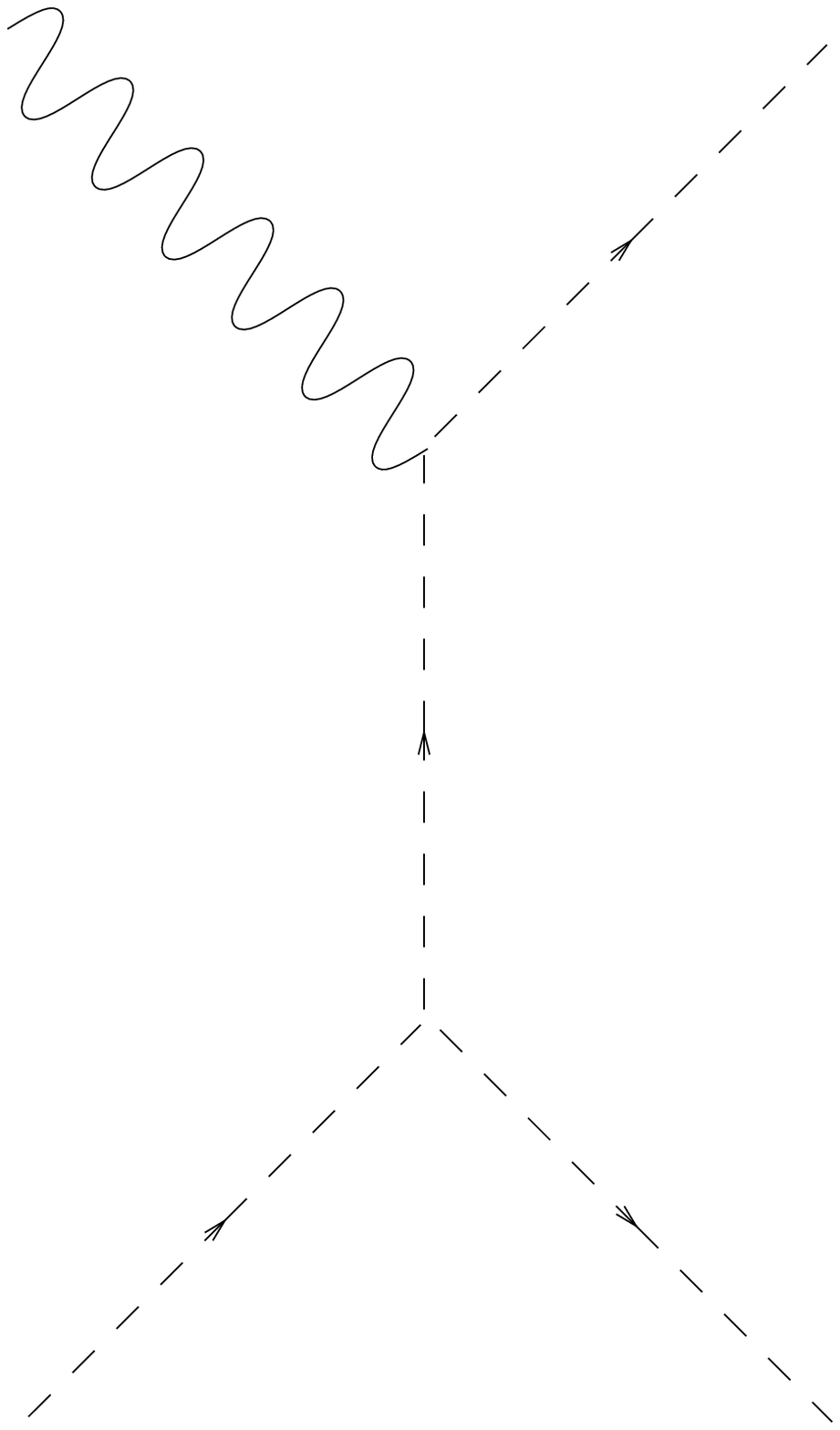,width=2truein,height=2truein}
 \caption{}
\end{center}
 \label{fig5}
 \end{figure}

 In this case,
 \be
 \Gamma_{Higgs}^{(2)}=\Sigma_{ijk}\frac{|\lambda_{ijk}|^2}{T}\int\frac{d^4k}{(2\pi )^4}
 \delta\left[(p+k)^2-m_j^2\right]\frac{Im
   \Pi (k)}{(k^2-m_k^2)^2}\coth(\frac{k_0}{2T}) 
 \ee
 Supposing that all the masses are $\sim T$, so that there is no IR
 divergence in the integral, and remembering that $Im
 \Pi \sim \alpha_{ew}T$, we obtain that 
 \be
 \Gamma_{Higgs}^{(2)}\sim \alpha_{ew}\Sigma_{ijk}\frac{|\lambda_{ijk}|^2}{T}
 \ee
 with a coefficient of order 1.
 \end{itemize}

 Putting everything together, we obtain the following transport
 equations:
 \beqa
 \label{transport_1}
 \dot q &=&
 D_{quark}\nabla^2q-\Gamma_{yukawa}\left(\frac{q}{6}-\frac{h}{8}-\frac{t}{3}
 \right)-\Gamma_{mass}\left(\frac{q}{6}-\frac{h_{eq}}{8}-\frac{t}{3}
 \right)\\
 \dot t &=&
 D_{quark}\nabla^2q+\Gamma_{yukawa}\left(\frac{q}{6}-\frac{h}{8}-\frac{t}{3}
 \right)+\Gamma_{mass}\left(\frac{q}{6}-\frac{h_{eq}}{8}-\frac{t}{3}
 \right)\\
 \label{transport_3}
 \dot h &=&
 D_{Higgs}\nabla^2h+\Gamma_{yukawa}\left(\frac{q}{6}-\frac{h}{8}-\frac{t}{3}
 \right)-\Gamma_{Higgs}\left(\frac{h-h_{eq}}{8}
 \right)
 \eeqa
 where $D_{quark}$, $D_{Higgs}$, $\Gamma_{yukawa}$,
 $\Gamma_{Higgs}\equiv\Gamma_{Higgs}^{(1)}+
 \Gamma_{Higgs}^{(2)}$ are given in Eqs.~(\ref{diff_quark}),
 (\ref{diff_Higgs}), (\ref{gamma_mass}), (\ref{yuk}),
 (\ref{gamma_Higgs}), 
(\ref{gamma_Higgs_2}).

 \section{Discussion and comparison with previous work}
\setcounter{equation}{0}
 Eqs.~(\ref{transport_1})-(\ref{transport_3}) were first derived in~{\cite{cohen336}}, 
 who interpreted ${\cal L}_{hyp}$ as an effective chemical potential 
 and used elementary
 thermodynamics. They made 
 reasonable assumptions to estimate 
 the $\Gamma$ and $D$ coefficients, 
  as well as both weak and
   strong sphaleron effects, without explicitly computing any of these
   quantities. In our approach, sphaleron contributions, which cannot be
   computed in perturbation   theory and have been ignored in all our
   discussion, have still to be 
   introduced by hand.  However, they only introduce additional terms 
 in Eqs.~(\ref{transport_1})-(\ref{transport_3}), without affecting 
the terms we  did  compute. 

 One might still wonder whether such a detailed computation is anything
 more than an academic exercise, given that there is no compelling
 evidence to support this specific model, let alone to constrain its many
 parameters. Nevertheless, in the course of our work we have reached a few
 conclusions that are worth pointing out:
 
 \begin {itemize}
 \item{{\bf Generalizations}: the technique we have used can be applied to any
     other model; admittedly, the thick wall approximation has simplified
     our task, but it can be relaxed: Eq.~({\ref{transport}}) is valid if
     the wall is thicker than $1/T$, a condition considerably
     weaker than the usual ``thick wall'' assumption, which requires $L_w
     >> \tau\sim 1/\alpha_s^2T$. If the latter constraint is not
     satisfied, but the first one is, we can still use
     Eq.~({\ref{transport}}), but we cannot assume local kinetic
     equilibrium. This means that the local densities are not simply
     parametrized by chemical potentials $\mu_{i}(x)$, and the whole
     computation becomes quite cumbersome, but hopefully still tractable.

 We should mention here that {\cite{nelson}} tried to avoid this
 complication, deriving a unified formalism that does not require the
 wall to be thin or thick. They derived transport equations similar to
 ours, by adding source terms (corresponding to our $h_{eq}$,$t_{eq}$,$q_{eq}$ 
 terms) to the unperturbed evolution. The  sources
 were computed for generic wall thickness, but taking only quadratic
 interactions into account; for this reason, they found vanishing effects
 in the thick wall limit, at odds with our own result. For realistic
 values of $L_w$ the numerical disagreement may not be important, but
 still it is not negligible.  
 }
 \item{{\bf Diffusion}: 
 The quark and Higgs diffusion constants that we obtained are noticeably smaller
 than the estimates in~{\cite{joyce281}}:
 $D_{quark}=6/T$,
 $D_{Higgs}=110/T$, whereas they are comparable with the values
 used  
 by {\cite{cohen336}} for their numerical analysis:  $D_{quark}=3/T$,
 $D_{Higgs}=10/T$;  for this reason we believe that the
 qualitative results and the discussion contained in {\cite{cohen336}}
 are still valid.
 }
 \item{{\bf Source terms:} in {\cite{cohen336}}, $\Gamma_{Higgs}$ and
     $\Gamma_{mass}$ were estimated to be:
 \be \label{est}
 \Gamma_{Higgs}\sim \Gamma_{mass}\sim \frac{m_t^2}{T}
 \ee
 As we said earlier, $\Gamma_{mass}$ is sensitive to the quark mean free
 path, and the value we obtained in Eq.~(\ref{gamma_mass}) is considerably
 smaller. $\Gamma_{Higgs}$ strongly depends on the unknown parameters of
 the Higgs sector; Eq.~(\ref{gamma_Higgs_2})  can be used to study the whole parameter space,
 but realistically we won't obtain results that are too different from
 ({\ref{est}}). 
 }
 \item{{\bf Baryon number}: {\cite{cohen336}} have integrated
 Eq.~(\ref{transport_1})-(\ref{transport_3})
     numerically, and found that values of $n_B/s$ consistent
    with experimental estimates can be obtained in this model by
    taking $\xi \sim 10^{-2}-10^{-3}$. We have not performed the same
    calculation with our values of the $\Gamma$'s and the $D$'s, but we
    believe that their conclusions would not be qualitatively
    altered. Therefore, this model should still be regarded as a viable
    candidate for electroweak baryogenesis.
}
\end{itemize}

\centerline{ {\bf Acknowledgements} }
It is a pleasure to thank Andrew Cohen for suggesting this project and
for his insightful assistance and patient encouragement. 
This work was supported in part under grant DE-FG02-91ER40676.



\newpage
\baselineskip18pt
\begin{thebibliography}{99}
\bibliographystyle{srt}

\bibitem{dimopoulos} V.A. \ Kuzmin, V.A. \ Rubakov, M.E. \ Shaposhnikov,
 {\em Phys. Lett.} {\bf B155 },
36 (1985).

\bibitem{cohen93} See for example A.G.\ Cohen, D.B.\ Kaplan, A.E.\ Nelson,
{\em Ann. Rev. Nucl. Part. Sci.} {\bf 43}, 27 (1993), and references therein.


\bibitem{sakharov} A.D.\ Sakharov, {\em JETP Lett} {\bf 6},
24 (1967).

\bibitem{tHooft} G.\ t'Hooft, 
{\em Phys. Rev. Lett.} {\bf 37},
  8 (1976); {\em Phys. Rev.} {\bf D14},
  3432 (1976).

\bibitem{gavela} M.B.\ Gavela {\em et al.}, 
{\em Nucl. Phys.} {\bf B430},
  345 (1994); {\em Nucl. Phys.} {\bf B430},
  382 (1994).


\bibitem{chou} K.\ Chou, Zh.\ Su, L.\ Yu, {\em Phys. Rep.} {\bf 118},
  1 (1985).

\bibitem{turok} N.\ Turok, J.\ Zadrozny,
{\em Phys. Rev. Lett.} {\bf 65},
  2331 (1990); {\em Nucl. Phys.} {\bf B358},
  471 (1991). A.G.\ Cohen, D.B.\ Kaplan, A.E.\ Nelson, 
{\em Phys. Lett.} {\bf B263},
  86 (1991). 


\bibitem{cline} J.M.\ Cline, K.\ Kainulainen, A.P.\ Vischer,
  hep-ph/9506284.


\bibitem{cohen336} A.G.\ Cohen, D.B.\ Kaplan, A.E.\ Nelson, 
{\em Phys. Lett.} {\bf B336},
  41 (1994).

\bibitem{landsman} N.P.\ Landsman, Ch.G.\ van Weert, 
{\em Phys. Rep.} {\bf 145},
  141 (1987).

\bibitem{lebedev} V.V.\ Lebedev, A.V.\ Smilga, 
{\em Ann. Phys.} {\bf 202},
  229 (1990).

\bibitem{nelson} P.\ Huet, A.E.\ Nelson, 
{\em Phys. Lett.} {\bf B355},
  229 (1995); {\em Phys. Rev.} {\bf D53},
  4578 (1996).




\bibitem{joyce281} M.\ Joyce, T.\ Prokopec, N.\ Turok, hep-ph/9410281.

\end{thebibliography}
\end{document}